# Characterisation of the plutonium isotopic composition of a sediment core from Palomares, Spain, by low-energy AMS and alpha- spectrometry

E. Chamizo[1], M.C. Jiménez –Ramos[1], S.M. Enamorado[1], M. García-León[2], R. García-Tenorio[3], J.L. Mas[4], P. Masqué[5], J. Merino[6], J.A. Sanchez-Cabeza[5,7]

[1] Centro Nacional de Aceleradores (CNA), Isla de la Cartuja, 41092 Seville, Spain.

[2] Departamento de Física Atómica, Molecular y Nuclear, Universidad de Sevilla, 41012 Sevilla, Spain.

[3] Departamento de Fisica Aplicada II, E.T.S.A, Universidad de Sevilla, 41012 Sevilla, Spain.

[4] Departamento de Física Aplicada I, Universidad de Sevilla, 41012 Sevilla, Spain.

[5] Institut de Ciència i Tecnologia Ambientals - Department de Física, Universitat Autònoma de Barcelona, 08193 Bellaterra, Spain.

[6] Amphos 21, Pg. de Rubí 29-31, E-08197 Valldoreix, Spain.

[7] International Atomic Energy Agency, Marine Environment Laboratories, MC-98000, Monaco.

*Abstract*

The measurement of plutonium isotopes, $^{239}$Pu and $^{240}$Pu, at 670 kV on the compact accelerator mass spectrometry (AMS) system at the *Centro Nacional de Aceleradores* (CNA) in Seville, Spain, is now a reality. In this work we present first Pu AMS results for environmental samples: a sediment core collected in a submarine canyon in the Mediterranean coast of the Spanish region of Palomares, affected by a nuclear accident in 1966. From the study of the $^{240}$Pu/$^{239}$Pu atomic ratio profile, showing on average levels lower than 11%, we confirm that the weapon-grade plutonium released on land during the accident, with a characteristic $^{240}$Pu/$^{239}$Pu atomic ratio of 5.8%, has found its way into the marine environment. A two-plutonium sources mixture model (Palomares and *fallout*) is used to elucidate the percentage of the plutonium coming from the accident. As a validation exercise of the Pu AMS measuring technique and in order to obtain the $^{238}$Pu/$^{(239+240)}$Pu activity ratios, samples were also studied by alpha-spectrometry (AS). The obtained AS $^{239+240}$Pu activity concentration results fit in with the AMS ones in a wide dynamic range, thus validating the AMS technique.

**Address correspondence to:**

Elena Chamizo Calvo

Centro Nacional de Aceleradores (CNA),   Phone: +34 954 46 05 53

Avda. Thomas Alva Edison 7, Isla de la Cartuja,   Fax:  +34 954 46 01 45

41092 Seville, Spain.   e-mail: elechacal@alum.us.es

## 1.- Introduction

On the 17th of February 1966, a surface of 223 ha belonging to the Cuevas de Almanzora municipality, on the Mediterranean Southeastern coast of Spain, was contaminated with plutonium from two 1.5 Mt (*TNT* eq.) thermonuclear bombs [1]. Since then, many efforts have been devoted to assess the actual impact of the so-called "Palomares accident". Recent studies characterising hot-particles have been carried out by non-destructive techniques such as electron-microscopy and μ-PIXE [2,3], and the contamination has been monitored by alpha-spectrometry (AS) and gamma spectroscopy [2]. Another are of investigation has been the land-to-sea transport of the contamination, revealed by some studies performed by AS on sediments, plankton, and seaweed samples from the Palomares coast [4,5,6,7,8]. However, a step forward in the study of the influence of the accident entails the identification of traces of weapon-grade plutonium in low-level environmental samples.

To achieve this, mass spectrometry (MS) techniques capable of quantifying the atomic ratio between the two major plutonium isotopes, $^{239}$Pu ($T_{1/2}$=24110 y) and $^{240}$Pu ($T_{1/2}$=6524 y), at environmental levels are necessary. This parameter informs unambiguously about the plutonium source: it is 5.8% for the plutonium released at the 1966 accident [1], and rises to about 18% for global fallout in the Northern Hemisphere [9]. Among the different well-validated MS techniques, one of the most competitive in terms of sensitivity, isobar and molecular suppression is accelerator mass spectrometry (AMS), as offers detection limits in the 10 µBq range without almost any matrix effect [10].

The first Pu AMS results on Palomares samples were obtained in 2006 making use of the 600 kV compact AMS system at the ETH/PSI Zürich. Specifically, a set of soil samples collected at different distances from the impact point of one of the two bombs. We established that even the samples with $^{239+240}$Pu activity inventories similar to the expected ones for fallout plutonium showed $^{240}$Pu/$^{239}$Pu atomic ratios lower than 11%, emphasizing the potential of this parameter to assess Pu in the environment [11].

Following this work, here we present the first Pu AMS results on a sediment core from the Palomares marine environment. The interest of this study is double. First, these are the first Pu environmental results obtained on the 1 MV AMS system at the *Centro Nacional de Aceleradores* (CNA) Seville, Spain; second, they provide new information about the presence of weapon-grade plutonium in the Mediterranean continental shelf.

**2. Sample and Analytical Procedures**

The studied sediment core (URE) was collected in the Aguas river submarine canyon in 1991 during a sampling cruise (Fig.1) [12]. The area has been shown to significantly accumulate the plutonium that had been transported to the sea via river flooding and airborne relocation [13]. The sediment core (L=20 cm) was sliced every 1.5 cm and samples were stored frozen in plastic bags until dried and powdered in the laboratory. Ten grams of each slice were spiked with 20 mBq of $^{242}$Pu and digested with a mixture of concentrated $HNO_3/H_2O_2$. The plutonium fraction was finally separated and purified using ion-chromatography resins TEVA [14]. 20% of this fraction was used for the AMS measurements and 80% for the AS determinations.

The AMS measurements were carried out using the 1 MV compact AMS at the CNA, Seville [15]. Briefly, the isotopes of interest - $^{239,240}$Pu and $^{242}$Pu, which was used as the normalisation isotope-, were extracted from the $Cs^+$ sputtering ion source to form $^{x}Pu^{16}O^-$ out of the commonly used solid matrix of $PuO_3$, $Fe_2O_3$ and Al powder. Next, these anions were stripped in Ar gas to $Pu^{3+}$ at the terminal of the accelerator, working at 670 kV, with about 11% of yield, and, in a final stage, counted from the total energy signal provided by a gas ionisation detector with a 30 nm thickness $Si_3N_{3.2}$ window. The three masses were injected into the detector in an automated way by adjusting the bouncer of the low-energy magnet, the terminal voltage, and the high-energy electrostatic deflector, so that every isotope had the same magnetic rigidity before and after the acceleration process. Counting times of about 5 s were dedicated to $^{242}$Pu, 10 s to $^{239}$Pu, and 20 s to $^{240}$Pu. On average, 20 minutes were devoted per analysis. The typical instrumental error of a measurement was 2% SDM.

The AS measurements were performed on an alpha spectrometer Alpha Analyst (Canberra), equipped with 450 mm$^2$ passivated implanted planar silicon detectors. The disks with the electrodeposits were measured at 1.5 mm distance from the detector. The final counting efficiency was 34% [16]. About 4 days of counting were dedicated per analysis in this case.

3.- Results and discussion

The $^{239+240}$Pu activity concentrations and the $^{240}$Pu/$^{239}$Pu atom ratios for the twin aliquots of the different slices of the URE sediment core measured by both AS and AMS are given in Table 1. There is an excellent agreement between both methods, as evidenced in Fig. 2, thus validating the AMS technique.

Results point to the presence of anomalously high concentrations of plutonium, as only a few samples show $^{239+240}$Pu activity concentrations similar to the reported ones in margin sediments from the western Mediterranean, which on average are lower than 1.5 mBq/g [17,18,19]. The same conclusion can be drawn from the $^{240}$Pu/$^{239}$Pu atomic ratios, displayed as well in Table 1, as they show significant deviations from the expected one for global fallout. Indeed, from these ratios we can calculate the fractions of the total $^{239+240}$Pu activity associated with the Palomares weapon-grade plutonium (P$_{wg}$) by using a two-sources mixing model [9]. The results, displayed in Table 1, clearly demonstrate that the accident is responsible for more than 40% of the plutonium accumulated in most core sections.

The distribution of the $^{239+240}$Pu activity concentration with depth shows that most of the plutonium is homogeneously distributed in the first 16 cm, except for three samples –at 2 and 5-6 cm depth- that produce two relative maxima. From the analysis of duplicate aliquots of the same sections (Table 1), these maxima may be explained by the presence of heterogeneities (hot particles). Therefore, the concentration profile seems to be a consequence of mixing processes (waves, storms, turbidity currents, bioturbation), precluding establishment of a deposition chronology of Pu in the area.

The $^{240}$Pu/$^{239}$Pu atomic ratio distribution with depth (Fig. 3) evidences the influence of the Palomares accident in the whole core, as in all samples this ratio is lower than that expected for fallout plutonium in those latitudes, which may be about 15% if we extrapolate the reported results in [11] for soils from Seville. The lowest ratios are found in the first 9 cm of the core, suggesting that from 47% to 95% of the total plutonium activity originated from the Palomares accident, whereas from 9 to 17 cm depth the influence of the accident is continuously decreasing, reaching its minimum in the deepest layer, with just 15% of the total plutonium coming from the accident. This is related to the differential degree of mixing affecting the sediments at a given depth. This trend is not so evident in the $^{238}$Pu/$^{239+240}$Pu activity ratios (Table 1), due to the relatively large associated uncertainties of $^{238}$Pu and to the narrow variability range between the two involved sources (about 2% for weapon-grade plutonium and 3% for fallout [1]). However, the four samples with the highest $^{239+240}$Pu activities (URE-2;5;6, first aliquots) show the lowest activity ratios, of about 1.7%, in agreement with the predominant source. Overall, the bomb-derived plutonium accounts for about 84% of the total plutonium activity in the studied sediment core.

**4.- Conclusions**

The comparison of the $^{239+240}$Pu activity concentrations on environmental samples measured by AMS and alpha spectrometry has provided a definitive proof of the applicability of the AMS technique at the CNA. Moreover, a new experimental and unambiguous confirmation of the land-to-sea transport of the weapon-grade plutonium from Palomares was given.

5.- Acknowledgements

The authors are really indebted to the whole AMS group from the CNA and to HVEE for the technical support. This work has been financed through the projects FIS2004-0495 of the Spanish Ministry of Science and Education, and RNM-419 of the *Junta de Andalucía.*

**Figure Captions**

Figure 1. Location of the URE sediment core.

Figure 2. Comparison between the $^{239+240}$Pu activity concentrations measured by AS and by AMS for the twin aliquots of the URE samples.

Figure 3. Distribution of the $^{239+240}$Pu activity concentrations (♦, in mBq/g, left axis) and of the $^{240}$Pu/$^{239}$Pu atomic ratios (▲, right axis) with sample depth in the URE core. The empty symbols refer to the second aliquots of the samples at 2, 5 and 6 cm depth.

**Table Captions**

Table 1. Plutonium by AMS and AS for twin aliquots of the URE sediment core. The $^{240}$Pu/$^{239}$Pu isotopic ratios are given in number of atoms; the $^{238}$Pu/$^{239+240}$Pu ratios in alpha activities. The parameter P$_{wg}$ refers to the percentage of the total activity associated to the weapon-grade plutonium [9]. The acronym N.M. stands for non-measured. The symbol * indicates $^{238}$Pu/$^{239+240}$Pu activity ratios that could not be measured due to the presence of traces of $^{228}$Th [14].

**Figure 1**

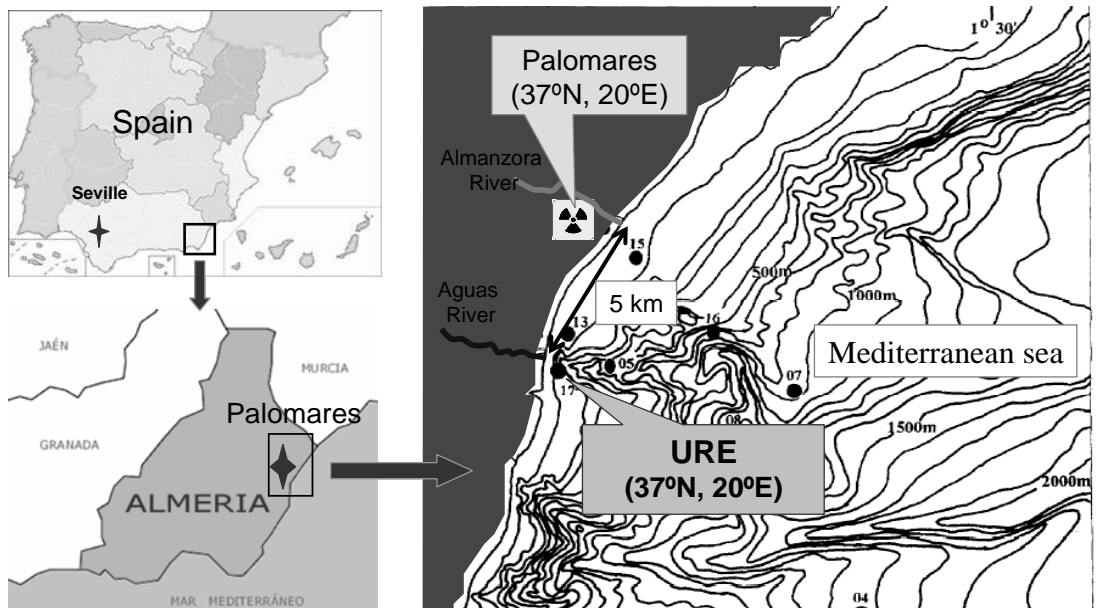

**Figure 2**

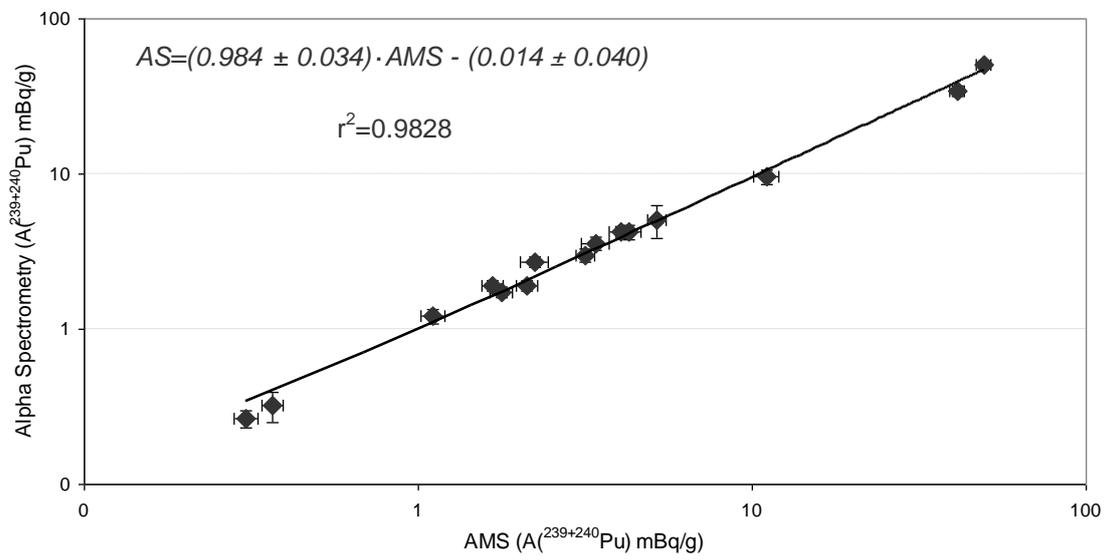

**Figure 3**

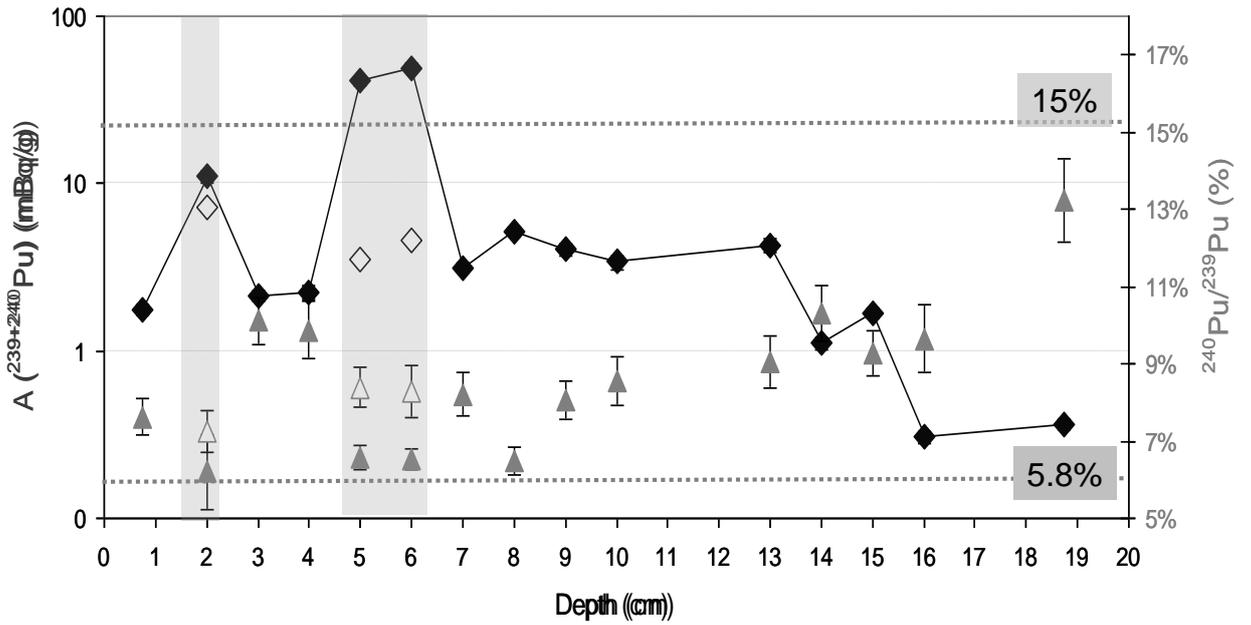

**Table 1**

| | | $^{239+240}$Pu (mBq/g) | | AMS | | AS |
|---|---|---|---|---|---|---|
| Sample | Layer (cm) | AMS | AS | $^{240}$Pu/$^{239}$Pu (%) | $P_{wg}$ (%)$^a$ | $^{238}$Pu/$^{239+240}$Pu |
| URE-1 | 0-1.5 | 1.78 ± 0.13 | 1.72 ± 0.13 | 7.63 ± 0.48 | 75.9 ± 7.7 | 0.0182 ± 0.0033 |
| URE-2 | 1.5-2.5 | 11.05 ± 0.99 | 9.55 ± 0.95 | 6.21 ± 0.97 | 94.4 ± 9.1 | 0.0152 ± 0.0037 |
| | | 7.19 ± 0.21 | N.M. | 7.27 ± 0.54 | 80.5 ± 7.8 | N.M. |
| URE-3 | 2.5-3.5 | 2.12 ± 0.15 | 1.91 ± 0.16 | 10.11 ± 0.62 | 47.1 ± 4.1 | 0.0239 ± 0.0047 |
| URE-4 | 3.5-4.5 | 2.24 ± 0.21 | 2.71 ± 0.21 | 9.88 ± 0.75 | 49.6 ± 4.5 | 0.0195 ± 0.0034 |
| URE-5 | 4.5-5.5 | 41.1 ± 2.2 | 34.3 ± 2.5 | 6.57 ± 0.30 | 89.5 ± 8.8 | 0.0160 ± 0.0014 |
| | | 3.552 ± 0.096 | N.M. | 8.38 ± 0.52 | 66.7 ± 5.6 | N.M. |
| URE-6 | 5.5-6.5 | 49.4 ± 2.6 | 50.4 ± 2.9 | 6.53 ± 0.27 | 90.1 ± 8.8 | 0.0170 ± 0.0011 |
| | | 4.59 ± 0.14 | N.M. | 8.28 ± 0.70 | 67.9 ± 6.3 | N.M. |
| URE-7 | 6.5-7.5 | 3.16 ± 0.20 | 3.01 ± 0.27 | 8.21 ± 0.56 | 68.8 ± 6.1 | 0.0215 ± 0.0047 |
| URE-8 | 7.5-8.5 | 5.18 ± 0.32 | 5.1 ± 1.2 | 6.48 ± 0.37 | 90.8 ± 9.7 | * |
| URE-9 | 8.5-9.5 | 4.05 ± 0.32 | 4.20 ± 0.35 | 8.07 ± 0.49 | 70.51 ± 6.1 | 0.0203 ± 0.0042 |
| URE-10 | 9.5-10.5 | 3.40 ± 0.33 | 3.57 ± 0.33 | 8.55 ± 0.63 | 64.7 ± 5.7 | 0.0254 ± 0.0051 |
| URE-12 | 12.5-13.5 | 4.28 ± 0.39 | 4.23 ± 0.44 | 9.05 ± 0.66 | 58.9 ± 5.2 | 0.0201 ± 0.0036 |
| URE-13 | 13.5-14.5 | 1.109 ± 0.090 | 1.21 ± 0.14 | 10.32 ± 0.72 | 44.7 ± 4.1 | 0.0289 ± 0.0079 |
| URE-14 | 14.5-15.5 | 1.67 ± 0.12 | 1.91 ± 0.14 | 9.26 ± 0.59 | 56.5 ± 4.8 | 0.0206 ± 0.0075 |
| URE-15 | 15.5-16.5 | 0.309 ± 0.021 | 0.269 ± 0.032 | 9.65 ± 0.86 | 52.1 ± 4.9 | 0.0216 ± 0.0083 |
| URE-17 | 16.5-19 | 0.367 ± 0.032 | 0.319 ± 0.071 | 13.2 ± 1.1 | 15.7 ± 2.2 | * |